\documentclass[aps,prl,twocolumn,showpacs,amsmath,amssymb]{revtex4-1}
\usepackage{epsfig}
\usepackage{graphicx}
\usepackage{dcolumn}
\usepackage{bm}
\usepackage{ltablex,booktabs}
\usepackage{overpic}
\usepackage{subfigure}
\usepackage{float}
\usepackage{color}
\usepackage{amsmath}
\usepackage{mathcomp}
\usepackage{mathrsfs}
\usepackage{multirow}
\usepackage{rotating}
\usepackage{amssymb}
\usepackage{gensymb}
\usepackage{amsmath}
\usepackage{tabularx}
\usepackage[bookmarksnumbered, pdfstartview=FitH,colorlinks,urlcolor=blue, citecolor=blue,linkcolor=blue] {hyperref}

\begin{document}
\normalsize
\parskip=5pt plus 1pt minus 1pt

\title{\boldmath Study of light scalar mesons through $D_s^+ \to \pi^0 \pi^0 e^+ \nu_e$ and $K_S^0 K_S^0 e^+ \nu_e$ decays}
\author{
\begin{small}
\begin{center}
M.~Ablikim$^{1}$, M.~N.~Achasov$^{10,b}$, P.~Adlarson$^{68}$, S. ~Ahmed$^{14}$, M.~Albrecht$^{4}$, R.~Aliberti$^{28}$, A.~Amoroso$^{67A,67C}$, M.~R.~An$^{32}$, Q.~An$^{64,50}$, X.~H.~Bai$^{58}$
% ...
\\
\vspace{0.2cm}
(BESIII Collaboration)\\
\vspace{0.2cm} {\it
$^{1}$ Institute of High Energy Physics, Beijing 100049, People's Republic of China\\
% ...
$^{4}$ Bochum Ruhr-University, D-44780 Bochum, Germany\\
% ...
$^{10}$ G.I. Budker Institute of Nuclear Physics SB RAS (BINP), Novosibirsk 630090, Russia\\
% ...
$^{14}$ Helmholtz Institute Mainz, Staudinger Weg 18, D-55099 Mainz, Germany\\
% ...
$^{28}$ Johannes Gutenberg University of Mainz, Johann-Joachim-Becher-Weg 45, D-55099 Mainz, Germany\\
% ...
$^{32}$ Liaoning Normal University, Dalian 116029, People's Republic of China\\
% ...
$^{50}$ State Key Laboratory of Particle Detection and Electronics, Beijing 100049, Hefei 230026, People's Republic of China\\
% ...
$^{58}$ University of Jinan, Jinan 250022, People's Republic of China\\
% ...
$^{64}$ University of Science and Technology of China, Hefei 230026, People's Republic of China\\
% ...
$^{67}$ University of Turin and INFN, (A)University of Turin, I-10125, Turin, Italy; (B)University of Eastern Piedmont, I-15121, Alessandria, Italy; (C)INFN, I-10125, Turin, Italy\\
$^{68}$ Uppsala University, Box 516, SE-75120 Uppsala, Sweden\\
% ...
$^{b}$ Also at the Novosibirsk State University, Novosibirsk, 630090, Russia\\
% ...
}
\end{center}
\vspace{0.4cm}
\end{small}
}
\noaffiliation{}

\date{\today}
\begin{abstract}
  Using 6.32~fb$^{-1}$ of $e^+e^-$ collision data recorded by the BESIII
  detector at center-of-mass energies between $4.178$ to $4.226$~GeV, we
  present the first measurement of the decay
  $D_s^+\to f_0(980)e^+\nu_e,\,f_0(980)\to \pi^0\pi^0$. The product branching
  fraction of $D_s^+\to f_0(980)e^+\nu_e,\,f_0(980)\to \pi^0\pi^0$ is measured
  to be $(7.9\pm1.4_{\rm stat} \pm0.4_{\rm syst})\times 10^{-4}$, with a
  statistical significance of $7.8\sigma$. Furthermore, the upper limits on the
  product branching fractions of $D_s^+\to f_0(500)e^+\nu_e$ with
  $f_0(500)\to \pi^0\pi^0$ and the branching fraction of
  $D_s^+\to K_{S}^{0}K_{S}^{0}e^+\nu_e$ are set to be $7.3\times 10^{-4}$ and
  $3.8\times 10^{-4}$ at 90\% confidence level, respectively. Our results
  provide valuable inputs to the understanding of the structures of light
  scalar mesons.
\end{abstract}
\maketitle

The constituent quark model has been strikingly successful, but the nontrivial
quark structures of scalar mesons below 1~GeV, $f_0(500)$, $f_0(980)$, and
$a_0(980)^{0(\pm)}$ (briefly denoted with $\sigma$, $f_0$, and $a_0^{0(\pm)}$,
respectively), are not completely classified~\cite{PDG}. Many theoretical
hypotheses, such as the tetraquark
states~\cite{Cheng:2017fkw, prd-15-267, prl-92-102001, plb-662-424, prd-79-074014, prd-97-036015, prl-110-261601, prd-99-014005, prd-103-014010, NPA-728-425, BCKa0, BCKa02},
and two-meson bound
states~\cite{prl-48-659, epja-37-303, ctp-58-410, plb-736-11, prd-92-034010},
have been proposed for these light scalar mesons but with controversial
results. Identifying the correct hypothesis is key to exploring 
chiral-symmetry-breaking mechanisms of non-perturbative QCD in low-energy
region~\cite{prd-15-267}. Therefore, conclusive experimental results are
required to interpret these states. 

Semileptonic charm meson decays provide a clean environment to study scalar
mesons~\cite{plb-759-501, prd-82-034016, prd-102-016022, prd-86-114010, IJMP-35-1460447, prd-92-054038, prd-102-016013}.
Experimentally, the BESIII collaboration has reported the measurements of
$D^{0(+)}\to a_0^{-(0)} e^+\nu_e$, and $D^+\to f_0/\sigma e^+\nu_e$ with
$f_0/\sigma \to \pi^+\pi^-$~\cite{prl-121-081802, prl-122-062001}, and the
search of $D_{s}^{+}\to a_0^{0} e^+\nu_e$~\cite{ref:a0980}. The CLEO
collaboration has also reported the measurement of $D_s^+\to f_0 e^+\nu_e$ with
$f_0 \to \pi^+\pi^-$~\cite{prd-80-052009}. On the other hand, theoretical
studies of neutral channels $(f_0 \to \pi^0\pi^0)$~ are rare compared to those
of charged channels. Like charged channels, the branching fractions~(BFs) of
the semileptonic $D^+_s$ decays into light scalar meson in their decay to
neutral channels and the $\pi^0\pi^0$ invariant mass spectrum aid in
understanding the nontrivial nature of light scalar
mesons~\cite{prd-86-114010, prd-102-016022, prd-103-014010, NPA-728-425}.
However, unlike charged channels, there is no background from
$\rho(770)^0\to\pi^+\pi^-$, thereby providing an ideal environment to study
$f_0/\sigma$. Therefore, it is of great interest to study this kind of decays
in experiment. 

In addition, the BaBar collaboration claimed that a possible $f_0\to K^+K^-$
contribution is found under the dominant decay $D_s^+\to \phi e^+ \nu_e$
in the study of $D_s^+\to K^+K^- e^+ \nu_e$~\cite{prd-78-051101}.
On the contrary, no other collaboration reported significant $f_0\to K^+K^-$
signal in the same decay~\cite{PDG}. We report the first search for the
neutral channel $D_s^+\to K_S^0K_S^0 e^+ \nu_e$, associated with
$f_0\to K_S^0K_S^0$, avoiding heavy contamination from $\phi\to K^+K^-$ decays.
Throughout this paper, charge conjugate channels are always implied.

The BESIII detector~\cite{Ablikim:2009aa, Ablikim:2019hff} records symmetric
$e^+e^-$ collisions provided by the BEPCII storage ring~\cite{Yu:IPAC2016-TUYA01}.
The cylindrical core of the BESIII detector covers 93\% of the full solid
angle and consists of a helium-based multilayer drift chamber~(MDC), a plastic
scintillator time-of-flight system~(TOF), and a CsI(Tl) electromagnetic 
calorimeter~(EMC), which are all enclosed in a superconducting solenoidal 
magnet providing a 1.0~T magnetic field. The charged-particle momentum
resolution at $1~{\rm GeV}/c$ is $0.5\%$, and the $dE/dx$ resolution is $6\%$
for electrons from Bhabha scattering. The EMC measures photon energies with a
resolution of $2.5\%$ ($5\%$) at $1$~GeV in the barrel (end cap) region. The
time resolution in the TOF barrel region is 68~ps, while that in the end cap
region is 110~ps. The end cap TOF system was upgraded in 2015 using multi-gap
resistive plate chamber technology, providing a time resolution of 60~ps~\cite{etof}.

The analysis is performed based on data samples corresponding to an integrated
luminosity of 6.32 fb$^{-1}$ at $\sqrt{s} = 4.178$, 4.189, 4.199, 4.209, 4.219,
and 4.226~GeV~\cite{ref:Kspipi0}. The signal events are selected from the
process $e^+e^-\to D_{s}^{*\pm}D_{s}^{\mp}\to \gamma D_{s}^{+}D_{s}^{-}$. A
{\sc geant4}-based~\cite{geant4} Monte Carlo (MC) simulation sample is used to
determine detection efficiencies and to estimate background processes. The
simulation models the beam energy spread and initial state radiation (ISR) in
the $e^+e^-$ annihilations with the generator {\sc kkmc}~\cite{ref:kkmc}. The
inclusive MC sample includes the production of open charm processes, the ISR
production of vector charmonium(-like) states, and the continuum processes
incorporated in {\sc kkmc}~\cite{ref:kkmc}. The known decay modes are modelled
with {\sc evtgen}~\cite{ref:evtgen} using BFs taken from the Particle Data
Group~\cite{PDG}, and the remaining unknown charmonium decays are modelled with
{\sc lundcharm}~\cite{ref:lundcharm}. Final state radiation~(FSR) from charged
final state particles is incorporated using {\sc photos}~\cite{photos}. The
signal detection efficiencies and signal shapes are obtained from signal MC
samples. In the signal MC sample, the $D_s^{-}$ decays generically and the
signal $D_s^{+}$ decays to $\pi^0\pi^0 e^{+} \nu_{e}$ or
$K_{S}^{0}K_{S}^{0} e^{+} \nu_{e}$ according to the generators described below.
The form factor $\mathcal{FF}$ is parameterized as~\cite{prd-46-5040, FF}
\begin{eqnarray} \begin{aligned}
  \mathcal{FF}=p_{\rm had}m_{D_{s}}\frac{\mathcal{A}}{1-\frac{q^2}{m_{A}^2}}\,,
\end{aligned}\end{eqnarray} 
where
$q^2$ is the invariant mass squared of $e^+\nu_e$ system, $p_{\rm had}$ is
magnitude of the three-momentum of the $\pi^0\pi^0/K_{S}^{0}K_{S}^{0}$ system
in the $D_s^+$ rest frame, the pole mass $m_{A}$ is expected to be
$m_{D_{s1}} \sim 2.5$~GeV/$c^{2}$~\cite{PDG}, and $m_{D_{s}}$ is the nominal
$D_{s}^{+}$ mass~\cite{PDG}. The amplitude $\mathcal{A}$ for the $f_{0}(980)$
resonance is parameterized by the Flatte formula with parameters fixed to the
LHCb measurement~\cite{flatte_f0_lhcb}, that for the $\sigma$ resonance is
described by the Bugg lineshape~\cite{ref:bugg}, and that in
$D_s^{+}\to K_{S}^{0}K_{S}^{0} e^{+} \nu_{e}$ signal MC sample is set to be
one.

The signal process
$e^{+}e^{-} \to D_{s}^{*+}D_{s}^{-}+c.c. \to \gamma D_{s}^{+}D_{s}^{-}+c.c$
allows studying semileptonic $D_{s}^{+}$ decays with a tag
technique~\cite{Li:2021iwf, MarkIII-tag} since the neutrino is the only one
particle undetected. There are two types of samples used in the tag technique:
single tag (ST) and double tag (DT). In the ST sample, a $D_{s}^{-}$ meson is
reconstructed through a particular hadronic decay without any requirement on
the remaining measured charged tracks and EMC showers. In the DT sample, a
$D_{s}^{-}$, designated as ``tag'', is reconstructed through a hadronic decay
mode first, and then a $D_{s}^{+}$, designated as the ``signal'', and the
transition photon from the $D_{s}^{*\pm}\to \gamma D_{s}^{\pm}$ decay are
reconstructed with the remaining tracks and EMC showers.
The BF of the signal decay is given by~\cite{ref:a0980}
\begin{eqnarray} \begin{aligned}
    \mathcal{B}_{\text{sig}}=\frac{N_{\text{total}}^{\text{DT}}}{\mathcal{B}_{\gamma}\sum_{\alpha,   
        i} N_{\alpha, i}^{\text{ST}}\epsilon^{\text{DT}}_{\alpha,
        i}/\epsilon_{\alpha, i}^{\text{ST}}},\, \label{eq:Bsig-gen}
\end{aligned} \end{eqnarray} 
where $\alpha$ represents various tag modes, $i$ denotes different $\sqrt{s}$,
$\epsilon_{\alpha, i}^{\rm DT (ST)}$ denotes the DT (ST) reconstruction
efficiencies, $\mathcal{B}_{\gamma}$ represents the BF of
$D_s^* \to\gamma D_s$, $N_{\text{total}}^{\text{DT}}$ is the signal yield for
all six data sets, and $N_{\alpha, i}^{\text{ST}}$ is the ST yields for various
tag modes. The tag candidates are reconstructed with charged $K$ and $\pi$,
$\pi^0$, $\eta^{(\prime)}$, and $K^0_S$ mesons in nine tag modes,
$D_s^-\to K_{S}^{0}K^{-}$, $K^{+}K^{-}\pi^{-}$, $K_{S}^{0}K^{-}\pi^{0}$,
$K^{+}K^{-}\pi^{-}\pi^{0}$, $K_{S}^{0}K^{-}\pi^{-}\pi^{+}$,
$K_{S}^{0}K^{+}\pi^{-}\pi^{-}$, $\pi^{-}\pi^{-}\pi^{+}$, $\pi^{-}\eta^{\prime}$,
and $K^{-}\pi^{+}\pi^{-}$. Requirements on the recoiling mass are applied to
the tag candidates in order to identify the process
$e^+e^-\to D^{*\pm}_sD^{\mp}_s$. If there are multiple candidates for a tag
mode, the one with recoiling mass closest to the nominal $D_s^{*\pm}$
mass~\cite{PDG} is chosen. A detailed description of the requirements on the
mass and the recoiling mass of tagged $D_s^{-}$, and the selection criteria for
charged and neutral particle candidates is provided in Ref.~\cite{ref:a0980}.
The ST yields of data for tag modes $N_{\alpha, i}^{\text{ST}}$ are determined
from fitting to the tag $D_{s}^{-}$ invariant mass~($M_{\rm tag}$)
distributions~\cite{SM}. The signal shape is modeled with the MC-simulated
shape convolved with a Gaussian function, and the background is parameterized
as a second-order Chebyshev function. The efficiencies $\epsilon$ for ST are
obtained from the inclusive MC samples~\cite{SM}.

\begin{figure}[htp]
  \begin{center}
    \includegraphics[width=0.235\textwidth]{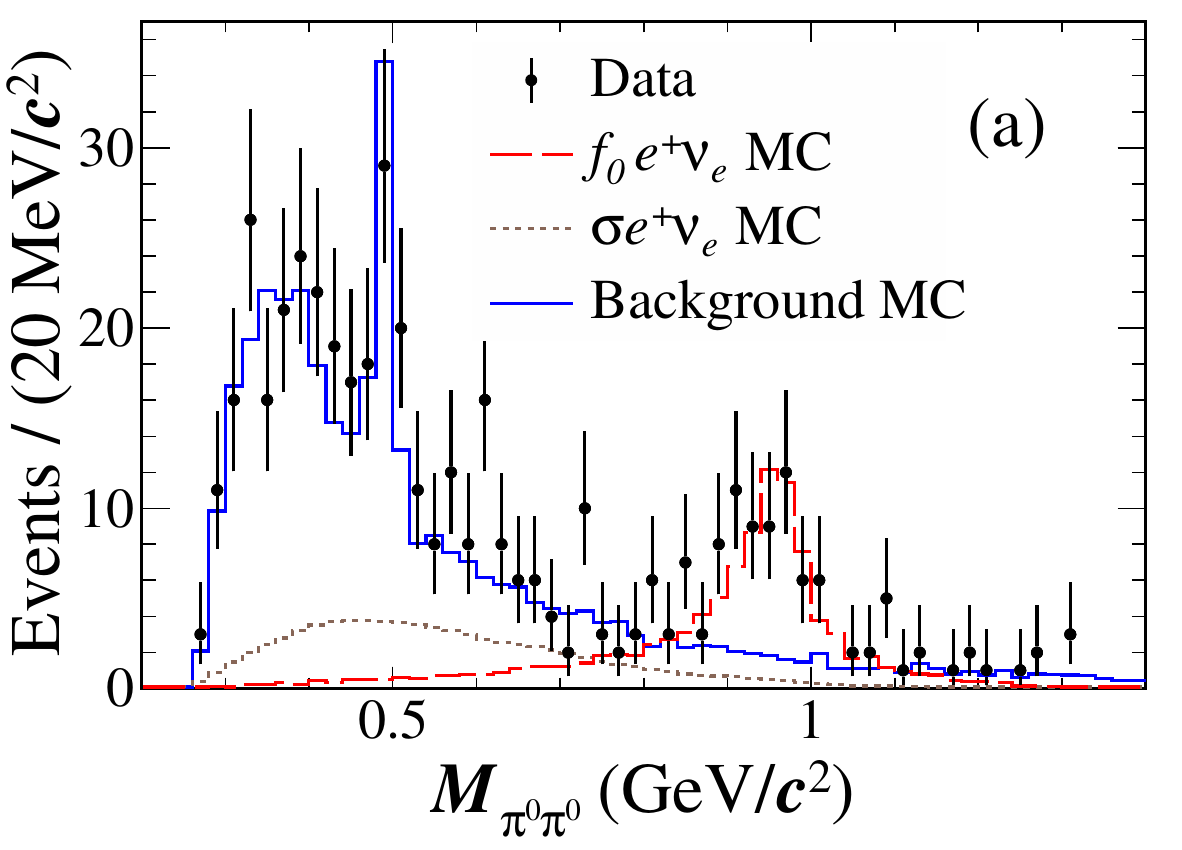}
    \includegraphics[width=0.235\textwidth]{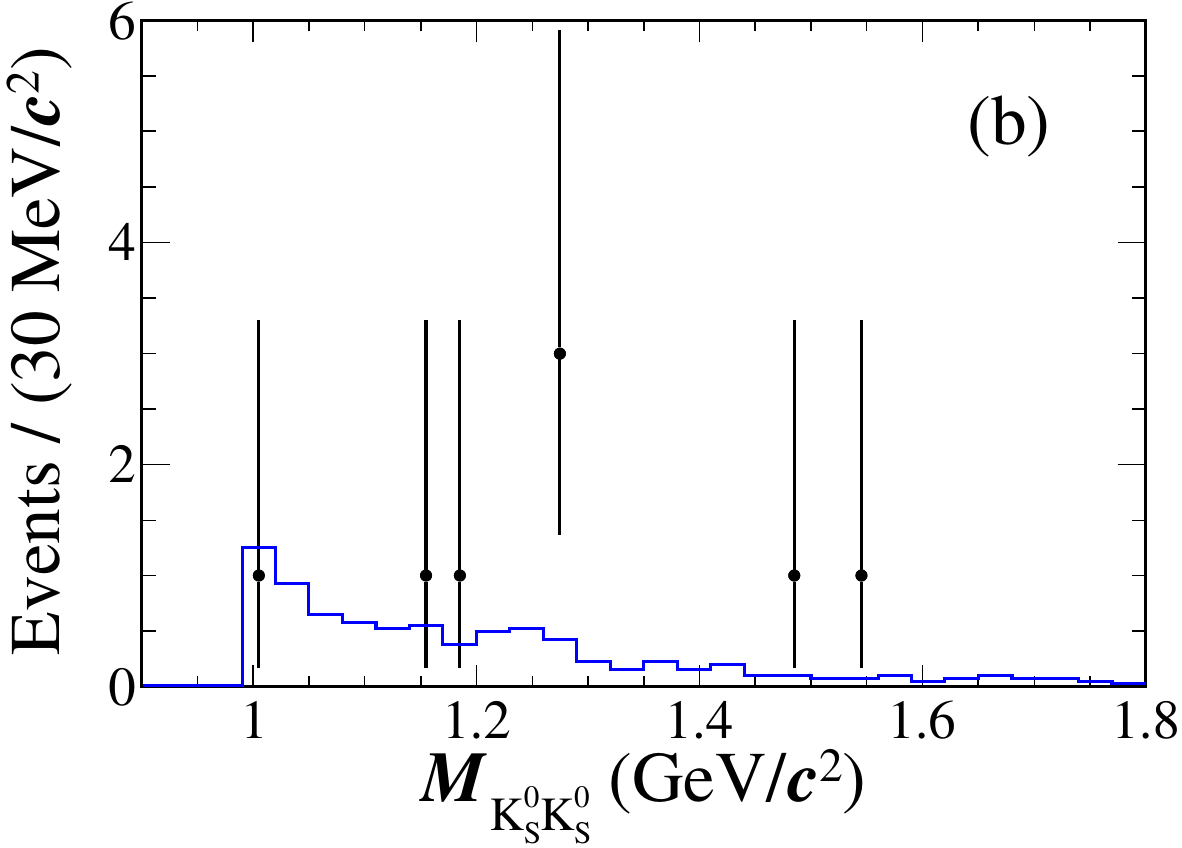}
    \caption{Invariant mass distributions of semileptonic-side hadrons of the
      selected candidates for (a) $D_{s}^{+}\to \pi^{0}\pi^{0}e^+\nu_e$ and (b)
      $D_{s}^{+}\to K_{S}^{0}K_{S}^{0}e^+\nu_e$. The points with error bars are
      data. The blue solid lines are the MC-simulated backgrounds. The peak
      around 0.5~GeV/$c^2$ in (a) is caused by the decay
      $D_{s}^{+}\to K^0_S(\to \pi^0\pi^0)e^+\nu_e$.
      The red dashed and brown dotted lines are signal MC samples of
      $D_{s}^{+}\to f_{0}(980)e^+\nu_e$ and $D_{s}^{+}\to \sigma e^+\nu_e$,
      respectively, which are normalized arbitrarily for visualization purposes.
      A cut on missing mass squared, $|{M\!M}^2|<0.15$~GeV$^2$/$c^4$, is applied.
    }
    \label{fig:hadron_mass}
  \end{center}
\end{figure}

After a tag $D_s^-$ is identified, the signal decays are selected recoiling
against the tag side, requiring that there is no track other than those
accounted for in the tagged $D_{s}^{-}$, the positron, and the
semileptonic-side hadrons ($N^{\text{extra}}_{\text{char}} = 0$). A joint
kinematic fit, in which four-momentum of the missing neutrino needs to be
determined, is performed to select the best transition photon candidate from
$D_s^{*\pm} \to \gamma D_s^{\pm}$. The fit includes: The total four-momentum of
reconstructed particles and the missing neutrino is constrained to the
four-momentum of $e^+e^-$ system. Invariant masses of the two $\pi^0/K_S^0$
candidates, the $D_s^-$ tag, the $D_s^+$ signal, and the $\gamma D_s^{\pm}$ are
constrained to the corresponding nominal masses~\cite{PDG}. The transition
photon candidate leading to the minimum $\chi^2$ of the joint kinematic fit is
chosen. Furthermore, the largest energy of the remaining EMC showers that are
not used to in the event reconstruction, $E^{\text{extra}}_{\gamma,\text{max}}$,
is required to be less than 0.2~GeV to suppress backgrounds with photon(s).
The square of the recoil mass against the transition photon and the $D_s^-$
tag~($M^{2}_{\rm rec}$) is expected to peak at the nominal $D^{\pm}_s$ meson
mass-squared before the kinematic fit for signal $D^{*\pm}_s D^{\mp}_s$ events.
Therefore, $M^{2}_{\rm rec}$ is required to satisfy
$3.75$~GeV$^{2}$/$c^4<M^{2}_{\rm rec}<4.05$~GeV$^{2}$/$c^4$ to suppress the
backgrounds from non-$D_sD_s^*$ processes. The missing neutrino is inferred by
the missing mass squared (${M\!M}^2$), defined as
\begin{eqnarray}
\begin{aligned}
{M\!M}^2=\frac{1}{c^2}(p_{\rm cm}-p_{\rm tag}-p_{\rm had}-p_{e}-p_{\gamma})^2\,,
\label{def:MM2}
\end{aligned}
\end{eqnarray}
where $p_{\rm cm}$ is the four-momentum of the $e^+e^-$ center-of-mass system,
$p_{\rm tag}$ for the tag $D_{s}^{-}$, $p_{\text{had}(e)}$ for the
semileptonic-side hadrons (positron), and $p_{\gamma}$ for the transition
photon from the $D_s^{*\pm}$ decay. To partially recover the energy lost due to
FSR and bremsstrahlung, the four-momenta of photon(s) within $5^{\circ}$ of the
initial positron direction are added to the positron four-momentum measured by
the MDC. The invariant mass distributions of semileptonic-side hadrons of the
selected candidates for $D_{s}^{+}\to \pi^0\pi^0 e^+\nu_e$ and
$D_{s}^{+}\to K_{S}^{0}K_{S}^{0}e^+\nu_e$ are  shown in
Fig.~\ref{fig:hadron_mass}. Notable $f_0$ signals are found in the $\pi^0\pi^0$
mass distribution while no significant signals of $\sigma\to \pi^0\pi^0$ and
$f_0\to K_{S}^{0}K_{S}^{0}$ are observed. The background is mostly caused by
miscellaneous backgrounds with multiple photons.

\begin{figure}[htp]
  \begin{center}
    \includegraphics[width=0.235\textwidth]{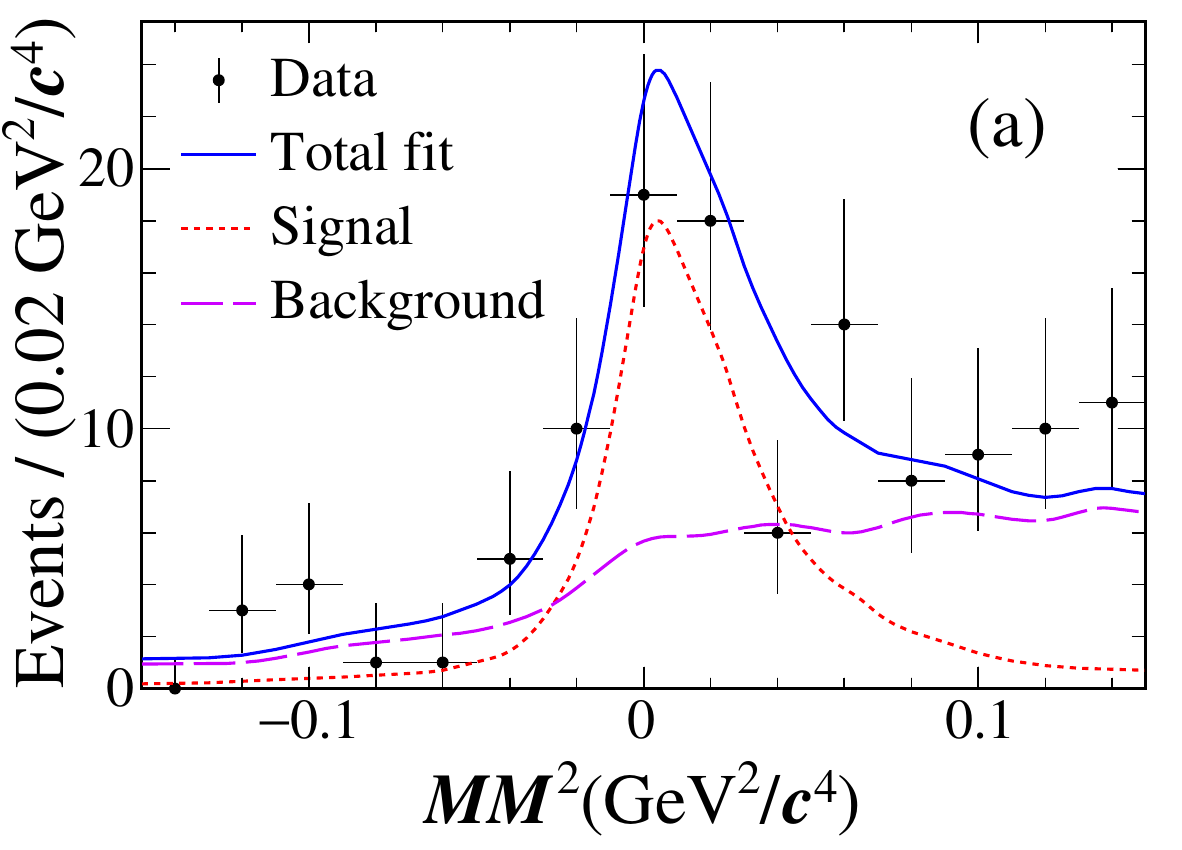}
    \includegraphics[width=0.235\textwidth]{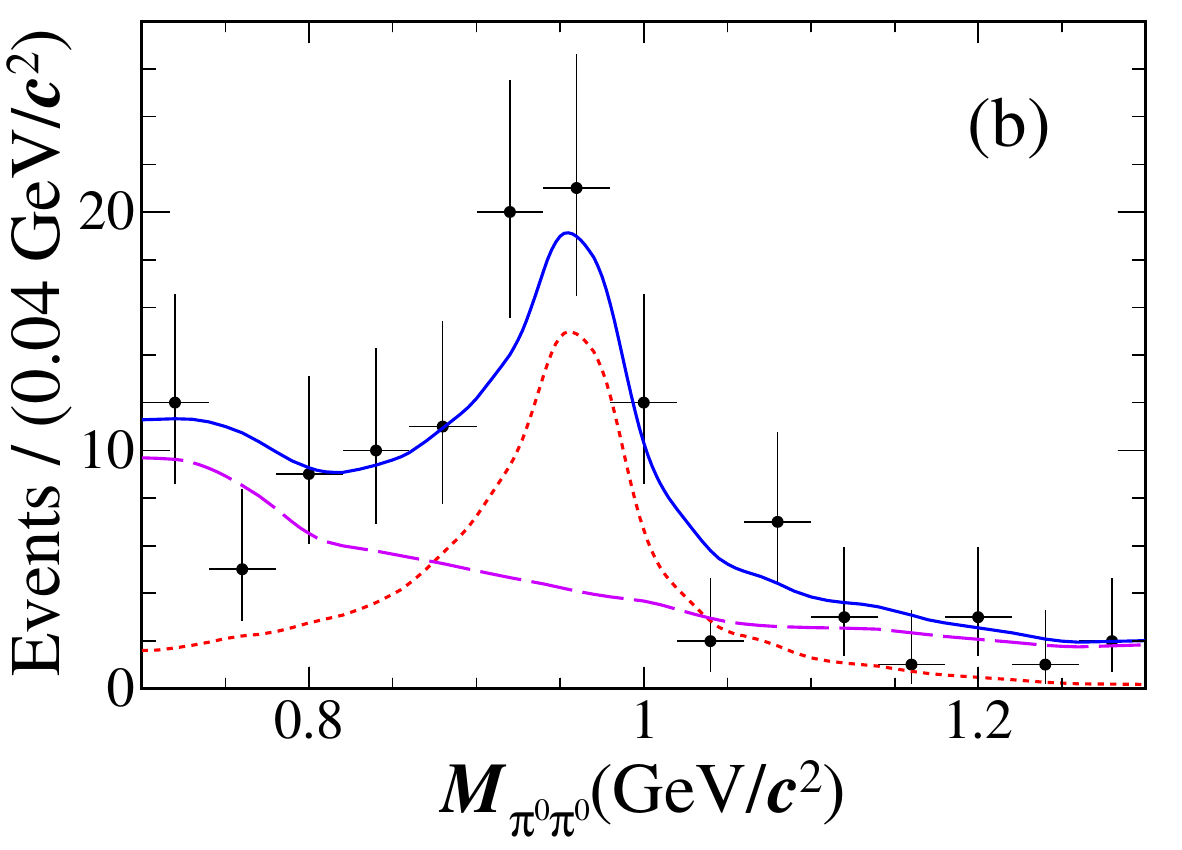}
    \caption{Projection on (a) $M\!M^2$ and (b) $M_{\pi^0\pi^0}$ of the
      two-dimensional fit to the selected candidates for
      $D_s^+\to \pi^0\pi^0e^+\nu_e$. The data are represented by points with
      error bars, the total fit result by blue solid lines, signal by red
      dashed lines, and background by violet long-dashed lines.}
    \label{fig:pi0_fit_data}
  \end{center}
\end{figure}

A two-dimensional unbinned maximum likelihood fit to the ${M\!M^2}$ versus
$M_{\pi^0\pi^0}$ distribution is performed to extract the DT yield of
$D_{s}^{+}\to f_0e^+\nu_e, f_0\to \pi^0\pi^0$. The signal and background
components are described by the simulated shape from the signal and inclusive
MC samples, respectively, using a kernel estimation method~\cite{kernel}
implemented in {\sc RooFit}~\cite{RooFit}. The fit result is shown in
Fig.~\ref{fig:pi0_fit_data}. The obtained signal yields is
$N_{\text{total}}^{\text{DT}}=54.8\pm10.1$ with a statistical significance of
$7.8\sigma$. Using the DT efficiencies from the signal MC
samples~(see Ref.~\cite{SM}), and $\mathcal{B}_{\gamma}$, the resulting
$\mathcal{B}(D_{s}^{+}\to f_0e^+\nu_e, f_0\to \pi^0\pi^0)$ is
$(7.9\pm1.4_{\rm stat} \pm 0.4_{\rm syst})\times 10^{-4}$. The second
uncertainty is systematic, which are described in the following.

\begin{figure}[htp]
  \begin{center}
    \includegraphics[width=0.235\textwidth]{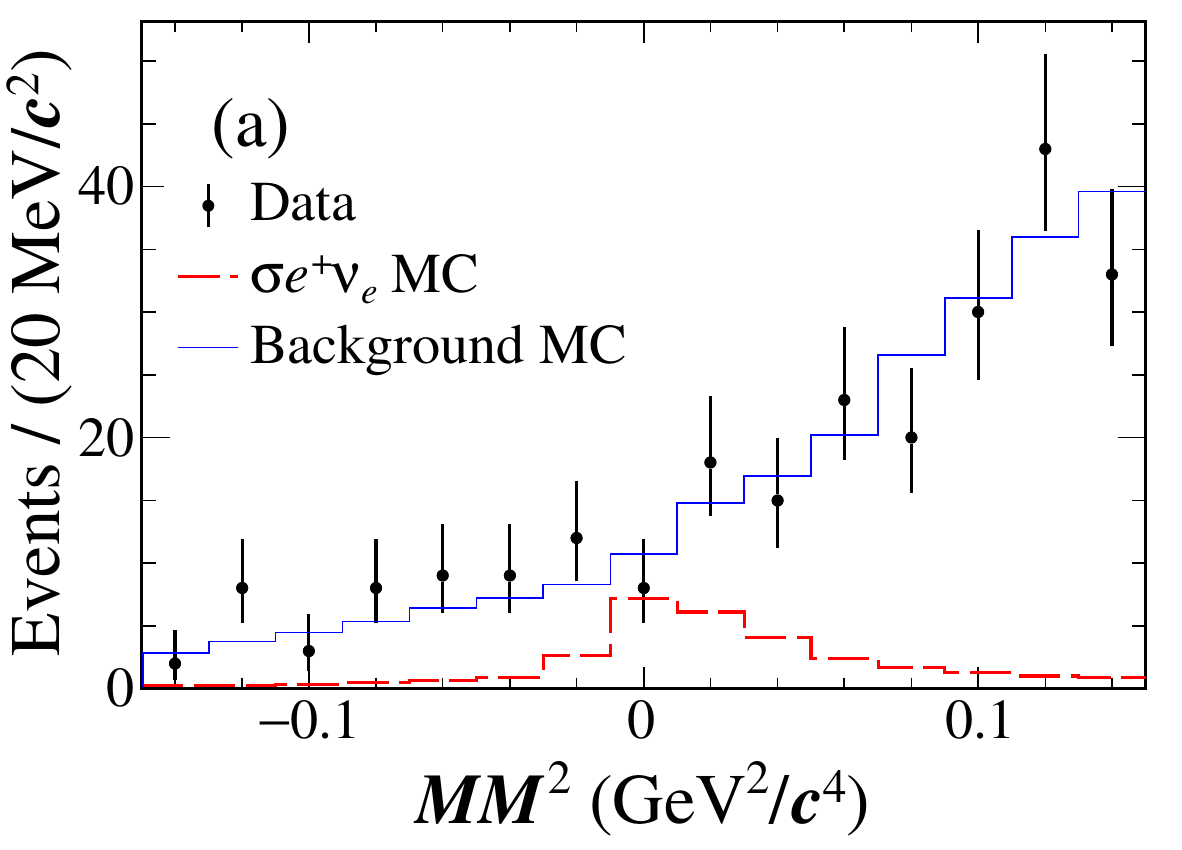}
    \includegraphics[width=0.235\textwidth]{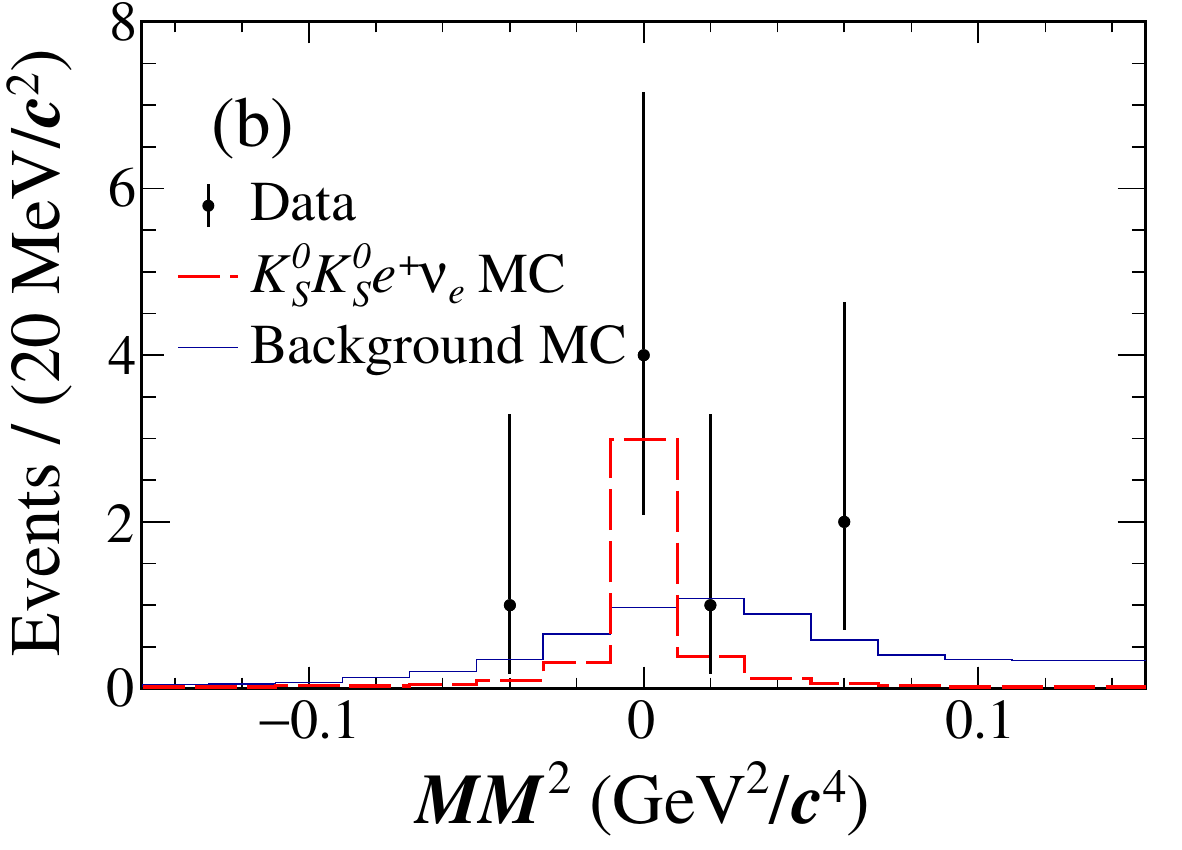}
    \includegraphics[width=0.235\textwidth]{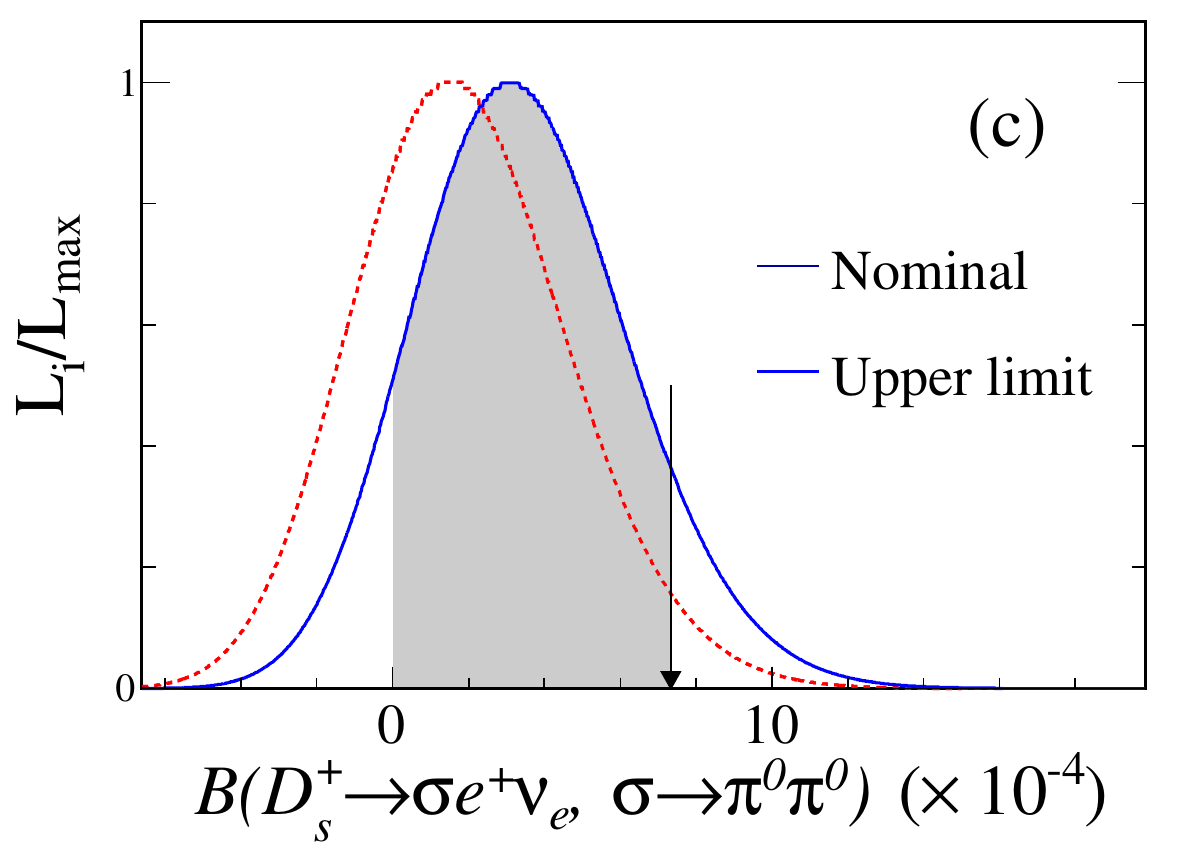}
    \includegraphics[width=0.235\textwidth]{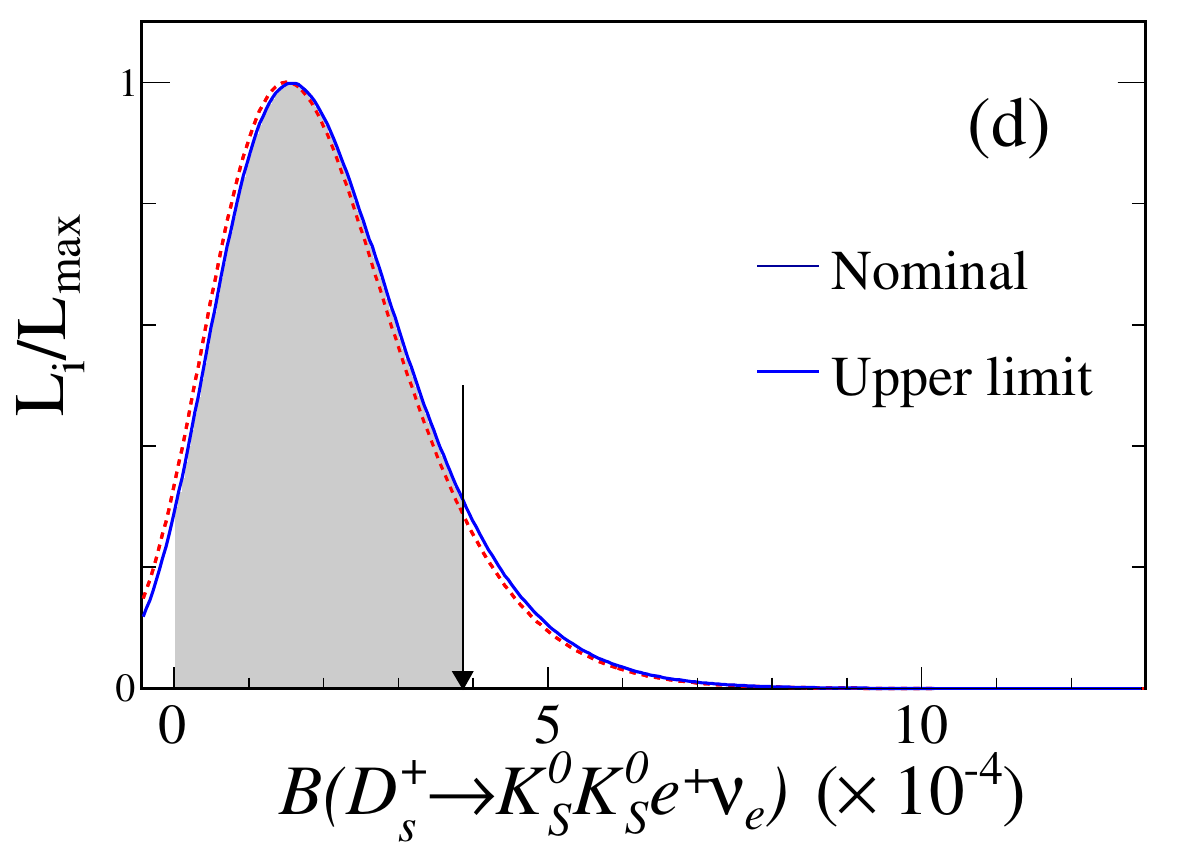}
    \caption{(top) ${M\!M^2}$ distributions and (bottom) likelihood
      distributions versus BF for (left)
      $D_{s}^{+}\to \sigma e^+\nu_e, \sigma\to \pi^0\pi^0$ and (right)
      $D_{s}^{+}\to K_{S}^{0}K_{S}^{0}e^+\nu_e$. The points with error bars are
      data, the blue solid lines are the MC-simulated backgrounds, and the red
      dashed lines show the MC-simulated signal shapes in (a, b). The signal
      shapes are normalized using an appropriate scaling factor chosen to
      visualize the shape and position of the signal. The red dashed lines in
      (c, d) are the likelihood curves for the nominal fit models, while the
      blue solid lines represent the likelihood curves that gives the upper
      limits after incorporating the systematic uncertainties. The black arrows
      indicate the results corresponding to 90\% C.L.
    }
\label{fig:UL_MM2}
\end{center}
\end{figure}

Since no significant signals are observed for the decays 
$D_{s}^{+}\to \sigma e^+\nu_e$ with $\sigma\to \pi^0\pi^0$ and 
$D_{s}^{+}\to K_S^0K_S^0e^+\nu_e$, the upper limits of the BFs for these decays
are determined. The candidate events for the former decay are required to
satisfy $M_{\pi^0\pi^0}< 0.66$~GeV/$c^2$. A veto
$0.458<M_{\pi^0\pi^0}<0.520$~GeV/$c^2$ is applied to suppress the background
from $D_{s}^{+}\to K^0_S(\to\pi^0\pi^0)e^+\nu_e$. Unbinned maximum-likelihood
fits are performed to the corresponding ${M\!M}^{2}$ distributions The signal
and background are modeled by the simulated shapes obtained from the signal and
inclusive MC samples, respectively. The ${M\!M^2}$ distributions and the
likelihoods of fit results as functions of assumed BFs are presented in
Fig.~\ref{fig:UL_MM2}. The upper limits, set at $90\%$ C.L., of the BFs of
$D_{s}^{+}\to \sigma e^+\nu_e, \sigma\to \pi^0\pi^0$ and
$D_s^+\to K_{S}^{0}K_{S}^{0}e^+\nu_e$ are $7.3 \times 10^{-4}$ and
$3.8 \times 10^{-4}$, respectively. The method to incorporate systematic
uncertainty is discussed in the following.

The sources of the systematic uncertainties for the BF measurement of
$D_{s}^{+}\to f_0e^+\nu_e$, as summarized in Table~\ref{tab:Sys}, are described
below. Note that most systematic uncertainties on the tag side cancel due to
the DT technique. Any residual effects are negligible. 

The uncertainty in the total number of the ST $D_s^-$ mesons is assigned to be
0.4\% by examining the changes of the fit yields when varying the signal shape,
background shape, and taking into account the background fluctuation in the
fit. The uncertainty from the quoted BF of $D^*_s\to \gamma D_s$ is
$0.7\%$~\cite{PDG}. The systematic uncertainties from tracking and PID
efficiencies of $e^+$ are assigned as $1.0\%$ for each by using radiative
Bhabha events. The systematic uncertainties from reconstruction efficiencies of
$\gamma$ and $\pi^0$ are studied by using control samples of the decay
$J/\psi\to\pi^{+}\pi^{-}\pi^{0}$~\cite{EPJC-76-369, CPC-40-113001} and the
process $e^+e^-\to K^+K^-\pi^+\pi^-\pi^0$, respectively. A conservative
2\%(1\%) systematic uncertainty is assigned for each
$\pi^0$(the transition photon) in the analysis of $D_s^+\to\sigma e^+\nu_e$,
since no significant signal is available to check the data-MC consistency. As
for the analysis of $D_s^+\to f_0e^+\nu_e$, a momentum-weighted correction
factor for each $\pi^0$ is calculated to be $99.4\%$ and the residual
uncertainty of 0.8\% is assigned as the corresponding systematic uncertainty
along with a 1\% systematic uncertainty for the transition photon. The
uncertainties of the $E^{\text{extra}}_{\gamma, \text{max}}<0.2$~GeV and
$N^{\text{extra}}_{\text{char}} = 0$ requirements are assigned as $0.7\%$ and 
$0.8\%$, respectively, by analyzing DT hadronic events of $\pi^\pm\pi^0\eta$. 
The uncertainty due to the limited MC statistics is obtained by
$\sqrt{\begin{matrix} \sum_{\alpha} (f_{\alpha}\frac{\delta_{\epsilon_{\alpha}}}{\epsilon_{\alpha}})^2\end{matrix}}$,
where $f_{\alpha}$ is the tag yield fraction in data, and $\epsilon_{\alpha}$ and 
$\delta_{\epsilon_{\alpha}}$ are the signal efficiency and the corresponding 
uncertainty of tag mode $\alpha$, respectively. The systematic uncertainty
associated with signal models is studied by replacing the parameters of $f_0$ from 
LHCb~\cite{flatte_f0_lhcb} by those from BES~\cite{flatte_f0} in generating
the signal MC sample. The difference of the measured BFs, where the effects of
the signal efficiencies and the two-dimensional signal shape have been taken
into account, is assigned as the associated systematic uncertainty.
The background shape is altered by varying the relative fractions of major
backgrounds from $e^+e^-\to q\bar{q}$ and non-$D_{s}^{*+}D_{s}^{-}$ open-charm
processes within 30\% according to the uncertainties of their input crossing section
in the inclusive MC sample. The effects caused by the smoothing parameter of the
kernel estimation method~\cite{kernel, RooFit} is negligible.
The largest change is taken as the corresponding systematic uncertainty. 

The sources of systematic uncertainties on the upper limit measurements are
classified into two types: additive~($\sigma_{n}$) and
multiplicative~($\sigma_{\epsilon}$).

Additive uncertainty is dominated by the background shape description. The
systematic uncertainty is studied by altering the nominal MC background shape
with two methods. First, alternative simulated shapes are used, where the
relative fractions of the dominant backgrounds from $e^+e^-\to q\bar q$ and
non-$D_{s}^{*\pm}D_{s}^{\mp}$ open-charm processes are varied within
  30\% according to the uncertainties of their input crossing section in the
  inclusive MC sample.
Second, the alternative background shapes are obtained from the
inclusive MC sample using the kernel estimation method~\cite{kernel, RooFit}
with the smoothing parameter varied to be 0, 1, and 2.

Multiplicative uncertainties, as summarized in Table~\ref{tab:Sys}, are related
to the efficiency determination and the quoted BFs. All systematic
uncertainties are the same as those for $D_{s}^{+}\to f_0e^+\nu_e$ except for 
the following. The uncertainty for the $K_{S}^{0}$ reconstruction efficiency is
assigned as $1.5\%$ per $K_{S}^{0}$ using control samples of 
$J/\psi\to K_{S}^{0}K^{\pm}\pi^{\mp}$ and $\phi K_{S}^{0}K^{\pm}\pi^{\mp}$
decays. The uncertainties of the $E^{\text{extra}}_{\gamma,\text{max}}<0.2$~GeV
and $N^{\text{extra}}_{\text{char}} = 0$ requirements in the 
$D_{s}^{+}\to K_{S}^{0}K_{S}^{0}e^+\nu_e$ study are assigned as $0.5\%$ and
$0.9\%$, respectively, by analyzing DT hadronic events of
$D^+_s\to K^+K^-\pi^{\pm}$ and $K_{S}^{0}K^{\pm}$. The systematic uncertainty of
the $\sigma$ modeling is considered by replacing the lineshape of $\sigma$ in
the signal MC sample with a conventional relativistic Breit-Wigner function with
the mass and width fixed to the BES measurements~\cite{plb-598-149}. The
systematic uncertainty related to the $K_S^0K_S^0 e^+ \nu_e$ model is estimated
by replacing the nominal model in the signal MC sample by a uniform
distribution in phase space.

The additive uncertainty is taken into account by extracting likelihood
distributions using different alternative background
shapes and the one resulting the most conservative upper limit is chosen.
Then, the multiplicative systematic uncertainty is incorporated in the
calculation of the upper limit via~\cite{Stenson, CPC-39-103001}
\begin{eqnarray}
\begin{aligned}
L\left(\mathcal{B}\right)\propto \int^1_0 L\left(\mathcal{B}\frac{\epsilon}{\epsilon_{0}}\right){\rm exp}\left[\frac{-\left(\epsilon/\epsilon_0-1\right)^2}{2 (\sigma_{\epsilon})^2 }\right]d\epsilon \,,
\end{aligned}
\end{eqnarray}
where $L(\mathcal{B})$ is the likelihood distribution as a function of BF;
$\epsilon$ is the expected efficiency and $\epsilon_0$ is the averaged
MC-estimated efficiency.

\begin{table}[htp]
  \begin{center}
    \caption{The systematic uncertainties (\%) in the BF measurements.
      Uncertainties associated with background shapes for $\sigma e^+\nu_e$ and
      $K_S^0K_S^0e^+\nu_e$ are additive in the upper limit measurements and not
      listed in this table.} 
    \vspace{0.50cm}
    \begin{tabular}{l|ccc}
      \hline\hline
      Source                                  & $f_0e^+\nu_e$  & $\sigma e^+\nu_e$  & $K_S^0K_S^0e^+\nu_e$\\
      \hline
      $D_{s}^{-}$ yield                       & 0.4            & 0.4                & 0.4\\
      ${\cal B}(D^{*\pm}\to \gamma D^{*\pm})$ & 0.7            & 0.7                & 0.7\\
      $e^+$ tracking efficiency               & 1.0            & 1.0                & 1.0\\
      $e^+$ PID efficiency                    & 1.0            & 1.0                & 1.0\\
      $\gamma$ and $\pi^0$ reconstruction     & 2.6            & 5.0                & 1.0\\
      $K_{S}^{0}$ reconstruction              &  -             &  -                 & 3.0\\
      $E^{\text{extra}}_{\gamma,\text{max}}<0.2$~GeV
                                              & 0.7            & 0.7                & 0.5\\
      $N^{\text{extra}}_{\text{char}} = 0$    & 0.8            & 0.8                & 0.9\\
      MC statistics                           & 0.5            & 0.5                & 0.5\\
      Signal model                            & 1.3            & 3.3                & 8.8\\
      Background shape                        & 3.0            & see text           &  see text \\
      \hline
      Total                                   & 4.7            & 6.3                & 9.5\\
      \hline\hline
    \end{tabular}
    \label{tab:Sys}
  \end{center}
\end{table}

In summary, the first BF measurement of
$D_s^+\to f_0 e^+\nu_e, f_0\to \pi^0\pi^0$ and searches for
$D_s^+\to \sigma e^+\nu_e, \sigma\to \pi^0\pi^0$ and
$D_s^+\to K^{0}_S K^{0}_S e^+\nu_e$ are performed using 6.32~fb$^{-1}$ of data
taken at $\sqrt{s} = 4.178-4.226$~GeV with the BESIII detector.

The BF of $D_s^+\to f_0 e^+\nu_e, f_0\to \pi^0\pi^0$ is determined to be
$(7.9\pm1.4_{\rm stat} \pm0.4_{\rm syst})\times 10^{-4}$.
According to isospin symmetry expectation
$\frac{\mathcal{B}(f_0\to\pi^0\pi^0)}{\mathcal{B}(f_0\to\pi^+\pi^-)}=0.5$, our
result is consistent with the measurement of $D_s^+\to f_0 e^+\nu_e$ with
$f_0 \to \pi^+\pi^-$ by the CLEO collaboration~\cite{prd-80-052009}. 
An upper limit on the BF of $D_s^+\to \sigma e^+\nu_e,\,\sigma\to \pi^0\pi^0$ is
set to be $7.3\times 10^{-4}$ at $90\%$ C.L. This upper limit is an
overestimation due to omitting the non-$\sigma$ contribution in the region of
$M_{\pi^0\pi^0}< 0.66$~GeV/$c^2$. Our results agree with the statement that the
$s\bar{s}\to \sigma$ transition is negligibly small in comparison with that of 
$s\bar{s}\to f_0$ given by Refs.~\cite{prd-86-114010, prd-92-054038}, which
follow the four-quark structure or meson–meson interaction hypothesis for $f_0$
and $\sigma$ mesons. Furthermore, the upper limit on
$\mathcal{B}(D_{s}^{+}\to K^{0}_S K^{0}_S e^+\nu_e)$ is set to be
$3.8\times 10^{-4}$ at $90\%$ C.L., indicating that contribution from
$\mathcal{B}(f_0\to K\bar{K})$ is not comparable to $\mathcal{B}(f_0\to \pi\pi)$
in semileptonic $D_s^+$ decays. Assuming $\mathcal{B}(f_0\to \pi^0\pi^0)$
contributes one third of the $f_0$ decays, our results leads to
$\mathcal{B}(D_{s}^{+}\to f_0 e^+\nu_e) = (2.4 \pm 0.4)\times 10^{-3}$, which
is consistent with the prediction given by
Refs.~\cite{prd-92-054038, prd-102-016013} when assuming $f_0$ to be the
admixture of $s\bar{s}$ and other light quark-antiquark pairs. 

\acknowledgements
The BESIII collaboration thanks the staff of BEPCII and the IHEP computing center for their strong support. This work is supported in part by National Key R\&D Program of China under Contracts Nos. 2020YFA0406400, 2020YFA0406300; National Natural Science Foundation of China (NSFC) under Contracts Nos. 11625523, 11635010, 11735014, 11822506, 11835012, 11875054, 11935015, 11935016, 11935018, 11961141012, 12022510, 12025502, 12035009, 12035013, 12061131003; the Chinese Academy of Sciences (CAS) Large-Scale Scientific Facility Program; Joint Large-Scale Scientific Facility Funds of the NSFC and CAS under Contracts Nos. U2032104, U1732263, U1832207; CAS Key Research Program of Frontier Sciences under Contract No. QYZDJ-SSW-SLH040; 100 Talents Program of CAS; INPAC and Shanghai Key Laboratory for Particle Physics and Cosmology; ERC under Contract No. 758462; European Union Horizon 2020 research and innovation programme under Contract No. Marie Sklodowska-Curie grant agreement No 894790; German Research Foundation DFG under Contracts Nos. 443159800, Collaborative Research Center CRC 1044, FOR 2359, FOR 2359, GRK 214; Istituto Nazionale di Fisica Nucleare, Italy; Ministry of Development of Turkey under Contract No. DPT2006K-120470; National Science and Technology fund; Olle Engkvist Foundation under Contract No. 200-0605; STFC (United Kingdom); The Knut and Alice Wallenberg Foundation (Sweden) under Contract No. 2016.0157; The Royal Society, UK under Contracts Nos. DH140054, DH160214; The Swedish Research Council; U. S. Department of Energy under Contracts Nos. DE-FG02-05ER41374, DE-SC-0012069.


\begin{thebibliography}{99}
\bibitem{PDG} P. A. Zyla {\it et al.} (Particle Data Group), \href{https://pdglive.lbl.gov/Viewer.action}{Prog. Theor. Exp. Phys. {\bf 2020}, 083C01 (2020).}
\bibitem{Cheng:2017fkw} X.~D.~Cheng, H.~B.~Li, B.~Wei, Y.~G.~Xu and M.~Z.~Yang, \href{https://journals.aps.org/prd/abstract/10.1103/PhysRevD.96.033002}{Phys.\ Rev.\ D {\bf 96}, 033002 (2017).}
\bibitem{prd-15-267} R. L. Jaffe, \href{https://journals.aps.org/prd/abstract/10.1103/PhysRevD.15.267}{Phys. Rev. D {\bf 15}, 267 (1977).}
\bibitem{NPA-728-425} N. N. Achasov, \href{https://www.sciencedirect.com/science/article/pii/S0375947403017378?via%3Dihub}{Nucl. Phys. A {\bf 728}, 425 (2003).}
\bibitem{prl-92-102001} J. R. Pel$\acute{\rm a}$ez, \href{https://journals.aps.org/prl/abstract/10.1103/PhysRevLett.92.102001}{Phys. Rev. Lett. {\bf 92}, 102001 (2004).}
\bibitem{plb-662-424} G.'t Hooft, G. Isidori, L. Maiani, A. D. Polosa, and V. Riquer, \href{https://www.sciencedirect.com/science/article/pii/S0370269308003651?via%3Dihub}{Phys. Lett. B {\bf 662}, 424 (2008).}
\bibitem{prd-79-074014} A. H. Fariborz, R. Jora, and J. Schechter, \href{https://journals.aps.org/prd/abstract/10.1103/PhysRevD.79.074014}{Phys. Rev. D {\bf 79}, 074014 (2009).}
\bibitem{prl-110-261601} S. Weinberg, \href{https://journals.aps.org/prl/abstract/10.1103/PhysRevLett.110.261601}{Phys. Rev. Lett. {\bf 110}, 261601 (2013).}
\bibitem{prd-97-036015} N. N. Achasov and A. V. Kiselev, \href{https://journals.aps.org/prd/abstract/10.1103/PhysRevD.97.036015}{Phys. Rev. D {\bf 97}, 036015 (2018).}
\bibitem{prd-99-014005} H. C. Kim, K. S. Kim, M. K Cheoun, D. Jido, and M. Oka, \href{https://journals.aps.org/prd/abstract/10.1103/PhysRevD.99.014005}{Phys. Rev. D {\bf 99}, 014005 (2019).}
\bibitem{prd-103-014010} N. N. Achasov, J. V. Bennett, A. V. Kiselev, E. A. Kozyrev, and G. N. Shestakov, \href{https://journals.aps.org/prd/abstract/10.1103/PhysRevD.103.014010}{Phys. Rev. D {\bf 103}, 014010 (2021).}
\bibitem{BCKa0} Y. K. Hsiao, Y. Yu and B. C. Ke, \href{https://doi.org/10.1140/epjc/s10052-020-08468-9}{Eur. Phys. J. C {\bf 80}, 895 (2020).}
\bibitem{BCKa02} Y.~Yu, Y.~K.~Hsiao, and B.~C.~Ke, \href{https://doi.org/10.1140/epjc/s10052-021-09895-y} {Eur. Phys. J. C \textbf{81}, 1093 (2021).}
\bibitem{prl-48-659} J. Weinstein and N. Isgur, \href{https://journals.aps.org/prl/pdf/10.1103/PhysRevLett.48.659}{Phys. Rev. Lett. {\bf 48}, 659 (1982).}
\bibitem{epja-37-303} T. Branz, T. Gutschea, and V. Lyubovitskij, \href{https://link.springer.com/article/10.1140%2Fepja%2Fi2008-10635-1}{Eur. Phys. J. A {\bf 37}, 303 (2008).}    
\bibitem{ctp-58-410} L. Y. Dai, X. G. Wang, and H. Q. Zheng, \href{https://iopscience.iop.org/article/10.1088/0253-6102/58/3/15}{Commun. Theor. Phys. {\bf 58}, 410 (2012).}
\bibitem{plb-736-11} L. Y. Dai and M. R. Pennington, \href{https://www.sciencedirect.com/science/article/pii/S0370269314004912?via%3Dihub}{Phys. Lett. B {\bf 736}, 11 (2014).}
\bibitem{prd-92-034010} T. Sekihara and S. Kumano, \href{https://journals.aps.org/prd/abstract/10.1103/PhysRevD.92.034010}{Phys. Rev. D {\bf 92}, 034010 (2015).}
\bibitem{prd-82-034016} W. Wang and C. D. Lu, \href{https://journals.aps.org/prd/abstract/10.1103/PhysRevD.82.034016}{Phys. Rev. D {\bf 82}, 034016 (2010).}
\bibitem{prd-86-114010} N. N. Achasov and A. V. Kiselev, \href{https://journals.aps.org/prd/abstract/10.1103/PhysRevD.86.114010}{Phys. Rev. D {\bf 86}, 114010 (2012).}
\bibitem{IJMP-35-1460447}  N. N. Achasov and A. V. Kiselev, \href{https://www.worldscientific.com/doi/epdf/10.1142/S2010194514604475}{Int. J. Mod. Phys. Conf. Ser. {\bf 35}, 1460447 (2014).}
\bibitem{plb-759-501} W. Wang, \href{https://www.sciencedirect.com/science/article/pii/S0370269316302581}{Phys. Lett. B {\bf 759}, 501 (2016).}
\bibitem{prd-92-054038}  T.~Sekihara and E.~Oset, \href{https://journals.aps.org/prd/abstract/10.1103/PhysRevD.92.054038}{Phys. Rev. D {\bf 92}, 054038 (2015).}
\bibitem{prd-102-016013}  N. R. Soni, A. N. Gadaria, J. J. Patel, and J. N. Pandya, \href{https://journals.aps.org/prd/abstract/10.1103/PhysRevD.102.016013}{Phys. Rev. D {\bf 102}, 016013 (2020).}
\bibitem{prd-102-016022} N. N. Achasov, A. V. Kiselev, and  G. N. Shestakov, \href{https://journals.aps.org/prd/abstract/10.1103/PhysRevD.102.016022}{Phys. Rev. D {\bf 102}, 016022 (2020).}
\bibitem{prl-121-081802} M. Ablikim {\it et al.} (BESIII Collaboration), \href{https://journals.aps.org/prl/abstract/10.1103/PhysRevLett.121.081802}{Phys. Rev. Lett. {\bf 121}, 081802 (2018).}
\bibitem{prl-122-062001} M. Ablikim {\it et al.} (BESIII Collaboration), \href{https://journals.aps.org/prl/abstract/10.1103/PhysRevLett.122.062001}{Phys. Rev. Lett. {\bf 122}, 062001 (2019).}
\bibitem{ref:a0980} M. Ablikim {\it et al.} (BESIII Collaboration), \href{https://journals.aps.org/prd/abstract/10.1103/PhysRevD.103.092004}{Phys. Rev. D {\bf 103}, 092004 (2021).}
\bibitem{prd-80-052009} K. M. Ecklund {\it et al.} (CLEO Collaboration), \href{https://journals.aps.org/prd/abstract/10.1103/PhysRevD.80.052009}{Phys. Rev. D {\bf 80}, 052009 (2009).}
\bibitem{prd-78-051101} B. Aubert {\it et al.} (BABAR Collaboration), \href{https://journals.aps.org/prd/abstract/10.1103/PhysRevD.78.051101}{Phys. Rev. D {\bf 78}, 051101(R) (2008).}
\bibitem{Ablikim:2009aa} M. Ablikim {\it et al.} (BESIII Collaboration), \href{https://www.sciencedirect.com/science/article/pii/S0168900209023870?via%3Dihub}{Nucl. Instrum. Methods Phys. Res. Sect. A {\bf 614}, 345 (2010).} 
\bibitem{Ablikim:2019hff} M. Ablikim {\it et al.} (BESIII Collaboration), \href{https://iopscience.iop.org/article/10.1088/1674-1137/44/4/040001}{Chin. Phys. C {\bf 44}, 040001 (2020).}
\bibitem{Yu:IPAC2016-TUYA01} C. H. Yu {\it et al.}, \href{https://inspirehep.net/files/96083dc6a03caf20597ac55b4500673a}{Proceedings of IPAC2016, Busan, Korea, 2016.}
\bibitem{etof} X.~Li {\it et al}., \href{https://link.springer.com/article/10.1007/s41605-017-0014-2}{Radiat. Detect. Technol. Methods {\bf 1}, 13 (2017);} Y.~X.~Guo {\it et al}., \href{https://link.springer.com/article/10.1007/s41605-017-0012-4}{Radiat. Detect. Technol. Methods {\bf 1}, 15 (2017).}
\bibitem{ref:Kspipi0} M. Ablikim {\it et al.} (BESIII Collaboration), \href{https://doi.org/10.1007/JHEP06(2021)181} {J. High Energ. Phys. {\bf 2021}, 181 (2021).}
\bibitem{geant4} S.~Agostinelli {\it et al.} (GEANT4 Collaboration), \href{https://inspirehep.net/files/6c9c0b62bbc8dc0401fca11a5fe5c87c}{Nucl. Instrum. Meth. A {\bf 506}, 250 (2003).}
\bibitem{ref:kkmc} S.~Jadach, B.~F.~L.~Ward and Z.~Was, \href{https://journals.aps.org/prd/abstract/10.1103/PhysRevD.63.113009}{Phys.\ Rev.\ D {\bf 63}, 113009 (2001);} \href{https://www.sciencedirect.com/science/article/pii/S0010465500000485?via%3Dihub}{Comput.\ Phys.\ Commun.\  {\bf 130}, 260 (2000).}
\bibitem{ref:evtgen} D.~J.~Lange, \href{https://www.sciencedirect.com/science/article/pii/S0168900201000894?via\%3Dihub}{Nucl.\ Instrum.\ Meth.\ A {\bf 462}, 152 (2001);} R.~G.~Ping, \href{https://iopscience.iop.org/article/10.1088/1674-1137/32/8/001}{Chin. Phys. C {\bf 32}, 599 (2008).}
\bibitem{ref:lundcharm} J.~C.~Chen, G.~S.~Huang, X.~R.~Qi, D.~H.~Zhang and Y.~S.~Zhu, \href{https://journals.aps.org/prd/pdf/10.1103/PhysRevD.62.034003}{Phys.\ Rev.\ D {\bf 62}, 034003 (2000);}
  R.~L.~Yang, R.~G.~Ping and H.~Chen, \href{https://iopscience.iop.org/article/10.1088/0256-307X/31/6/061301/pdf}{Chin.\ Phys.\ Lett.\  {\bf 31}, 061301 (2014).}
\bibitem{photos} E.~Richter-Was, \href{https://www.sciencedirect.com/science/article/pii/037026939390062M?via\%3Dihub}{Phys.\ Lett.\ B {\bf 303}, 163 (1993).}
\bibitem{FF} P. del Amo Sanchez {\it et al.} (BABAR Collaboration), \href{https://journals.aps.org/prd/abstract/10.1103/PhysRevD.83.072001}{Phys. Rev. D {\bf 83}, 072001 (2011).}
\bibitem{prd-46-5040} C. L. Y. Lee, M. Lu, and M. B. Wise, \href{https://lib-extopc.kek.jp/preprints/PDF/1992/9206/9206173.pdf}{Phys. Rev. D {\bf 46}, 5040 (1992).}
\bibitem{flatte_f0_lhcb} R. Aaij {\it et al.} (LHCb Collaboration), \href{https://journals.aps.org/prd/abstract/10.1103/PhysRevD.86.052006}{Phys. Rev. D {\bf 86}, 052006 (2012).}
\bibitem{ref:bugg} D. V. Bugg, \href{https://iopscience.iop.org/article/10.1088/0954-3899/34/1/011}{J. Phys. G {\bf 34}, 151 (2006).}
\bibitem{Li:2021iwf} H.~B.~Li and X.~R.~Lyu, \href{https://academic.oup.com/nsr/article/8/11/nwab181/6381732}{Natl. Sci. Rev. {\bf 8}, no. 11, nwab181 (2021).}
\bibitem{MarkIII-tag} J. Adler {\it et al.} (MARK-III Collaboration), \href{https://lib-extopc.kek.jp/preprints/PDF/1989/8903/8903163.pdf}{Phys. Rev. Lett. {\bf 62}, 1821 (1989).}  
\bibitem{SM} See Supplemental Material at [URL] for additional analysis information.
\bibitem{kernel} K. S. Cranmer, \href{https://www.sciencedirect.com/science/article/pii/S0010465500002435?via%3Dihub}{Comput. Phys. Commun. {\bf 136}, 198 (2001).}
\bibitem{RooFit} R. Brun and F. Rademakers, \href{https://www.sciencedirect.com/science/article/pii/S016890029700048X?via\%3Dihub}{Nucl. Instrum. Methods Phys. Res., Sect. A {\bf 389}, 81 (1997).}

\bibitem{EPJC-76-369} M. Ablikim {\it et al.} (BESIII Collaboration), \href{https://link.springer.com/article/10.1140%2Fepjc%2Fs10052-016-4198-2}{Eur. Phys. J. C {\bf 76}, 369 (2016).}
\bibitem{CPC-40-113001} M. Ablikim {\it et al.} (BESIII Collaboration), \href{https://iopscience.iop.org/article/10.1088/1674-1137/40/11/113001}{Chin. Phys. C {\bf 40}, 113001 (2016).}
\bibitem{flatte_f0} M. Ablikim {\it et al.} (BES Collaboration), \href{https://linkinghub.elsevier.com/retrieve/pii/S0370269304017265}{Phys. Lett. B {\bf 607}, 243 (2005).}
\bibitem{plb-598-149} M. Ablikim {\it et al.} (BES Collaboration), \href{https://www.sciencedirect.com/science/article/pii/S0370269304011384}{Phys. Lett. B {\bf 598}, 149 (2004).}
\bibitem{Stenson} K. Stenson, \href{https://arxiv.org/abs/physics/0605236}{arXiv:0605236[physics].}
\bibitem{CPC-39-103001} X. X. Liu, X. R. Lyu, and Y. S. Zhu, \href{https://iopscience.iop.org/article/10.1088/1674-1137/39/10/103001}{Chin. Phys. C {\bf 39}, 113001 (2015).}  
\end{thebibliography}
\end{document}


\normalsize
\parskip=5pt plus 1pt minus 1pt
\title{\boldmath Supplemental Material for ``Study of light scalar mesons through $D_s^+ \to \pi^0 \pi^0 e^+ \nu_e$ and $K_S^0 K_S^0 e^+ \nu_e$ decays''}
\author{
\begin{small}
\begin{center}
M.~Ablikim$^{1}$, M.~N.~Achasov$^{10,b}$, P.~Adlarson$^{68}$, S. ~Ahmed$^{14}$, M.~Albrecht$^{4}$, R.~Aliberti$^{28}$, A.~Amoroso$^{67A,67C}$, M.~R.~An$^{32}$, Q.~An$^{64,50}$, X.~H.~Bai$^{58}$
% ...
\\
\vspace{0.2cm}
(BESIII Collaboration)\\
\vspace{0.2cm} {\it
$^{1}$ Institute of High Energy Physics, Beijing 100049, People's Republic of China\\
% ...
$^{4}$ Bochum Ruhr-University, D-44780 Bochum, Germany\\
% ...
$^{10}$ G.I. Budker Institute of Nuclear Physics SB RAS (BINP), Novosibirsk 630090, Russia\\
% ...
$^{14}$ Helmholtz Institute Mainz, Staudinger Weg 18, D-55099 Mainz, Germany\\
% ...
$^{28}$ Johannes Gutenberg University of Mainz, Johann-Joachim-Becher-Weg 45, D-55099 Mainz, Germany\\
% ...
$^{32}$ Liaoning Normal University, Dalian 116029, People's Republic of China\\
% ...
$^{50}$ State Key Laboratory of Particle Detection and Electronics, Beijing 100049, Hefei 230026, People's Republic of China\\
% ...
$^{58}$ University of Jinan, Jinan 250022, People's Republic of China\\
% ...
$^{64}$ University of Science and Technology of China, Hefei 230026, People's Republic of China\\
% ...
$^{67}$ University of Turin and INFN, (A)University of Turin, I-10125, Turin, Italy; (B)University of Eastern Piedmont, I-15121, Alessandria, Italy; (C)INFN, I-10125, Turin, Italy\\
$^{68}$ Uppsala University, Box 516, SE-75120 Uppsala, Sweden\\
% ...
$^{b}$ Also at the Novosibirsk State University, Novosibirsk, 630090, Russia\\
% ...
}
\end{center}
\vspace{0.4cm}
\end{small}
}
\noaffiliation{}

\maketitle




The data samples used in this analysis are collected at $\sqrt{s} = 4.178$, 4.189, 4.199, 4.209, 4.219, 4.226~GeV.
For some aspects of the analysis, these samples are organized into three
sample groups, 4.178~GeV, 4.189-4.219~GeV, and 4.226~GeV, that were acquired
during the same year under consistent running conditions.

The tag $D_s^-$ invariant mass $M_{\rm tag}$ requirement, ST tag yield, and ST efficiency for each tag mode are
listed in Table~\ref{tab:tag-eff}.
The DT efficiencies for $D_s^{+}\to f_0(980)e^+\nu_e$ with $f_0(980)\to\pi^0\pi^0$,
$D_s^{+}\to f_0(500)e^+\nu_e$ with $f_0(500)\to\pi^0\pi^0$, and
$D_s^{+}\to K_S^0K_S^0 e^+ \nu_e$ are listed in Tables~\ref{tab:pi0_dtagEff2},~\ref{tab:sigma_dtagEff2},~and~\ref{tab:KsKs_dtagEff2}, respectively.

\begin{table*}[htbp]
  \renewcommand\arraystretch{1.25}
  \caption{ST yields~($N_{\alpha, i}^{\rm ST}$) and ST efficiencies~($\epsilon_{\alpha, i}^{\rm ST}$)
    for (I) $\sqrt{s} = 4.178$~GeV, (II) $4.189-4.219$~GeV, and (III) $4.226$~GeV. Uncertainties are
    statistical only. These efficiencies do not include the following intermediate-state branching
    fractions for $K_S^0\to \pi^+\pi^-$, $\pi^0\to\gamma\gamma$, $\eta\to \gamma\gamma$, and
    $\eta^{\prime}\to\pi^+\pi^-\eta$. 
    }\label{tab:tag-eff}
    \begin{center}
      \begin{tabular}{lccc}
        \hline
        \hline
        Tag mode                                     & $M_{\rm tag}$ (GeV/$c^{2}$)  & (I) $N_{\alpha, i}^{\rm ST}$       & (I) $\epsilon_{\alpha, i}^{\rm ST}$ \\
        \hline
        $D_{s}^{-} \to K_{S}^{0}K^{-}$               & [1.948, 1.991]             & $\phantom{0}31941 \pm 312$   & $47.36 \pm 0.07$\\
        $D_{s}^{-} \to K^{+}K^{-}\pi^{-}$            & [1.950, 1.986]             & $137240           \pm 614$   & $39.47 \pm 0.03$\\
        $D_{s}^{-} \to K_{S}^{0}K^{-}\pi^{0}$        & [1.946, 1.987]             & $\phantom{0}11385 \pm 529$   & $16.12 \pm 0.11$\\
        $D_{s}^{-} \to K^{+}K^{-}\pi^{-}\pi^{0}$     & [1.947, 1.982]             & $\phantom{0}39306 \pm 799$   & $10.50 \pm 0.03$\\
        $D_{s}^{-} \to K_{S}^{0}K^{-}\pi^{-}\pi^{+}$ & [1.958, 1.980]             & $\phantom{00}8093 \pm 326$   & $20.40 \pm 0.12$\\
        $D_{s}^{-} \to K_{S}^{0}K^{+}\pi^{-}\pi^{-}$ & [1.953, 1.983]             & $\phantom{0}15719 \pm 289$   & $21.83 \pm 0.06$\\
        $D_{s}^{-} \to \pi^{-}\pi^{-}\pi^{+}$        & [1.952, 1.982]             & $\phantom{0}37977 \pm 859$   & $51.43 \pm 0.15$\\
        $D_{s}^{-} \to \pi^{-}\eta^{'}$              & [1.940, 1.996]             & $\phantom{00}7759 \pm 141$   & $19.12 \pm 0.06$\\
        $D_{s}^{-} \to K^{-}\pi^{+}\pi^{-}$          & [1.953, 1.986]             & $\phantom{0}17423 \pm 666$   & $47.46 \pm 0.22$\\
        \hline
        \hline
      \end{tabular}
      \begin{tabular}{lcccc}
        \hline\hline
        Tag mode  & (II) $N_{\alpha, i}^{\rm ST}$ & (II) $\epsilon_{\alpha, i}^{\rm ST}$ & (III) $N_{\alpha, i}^{\rm ST}$ & (III) $\epsilon_{\alpha, i}^{\rm ST}$\\
        \hline
        $D_{s}^{-} \to K_{S}^{0}K^{-}$               & $18559           \pm 261$ & $47.26 \pm 0.09$         & $\phantom{0}6582 \pm 160$  & $46.37 \pm 0.16$\\
        $D_{s}^{-} \to K^{+}K^{-}\pi^{-}$            & $81286           \pm 505$ & $39.32 \pm 0.04$         & $28439           \pm 327$  & $38.38 \pm 0.07$\\
        $D_{s}^{-} \to K_{S}^{0}K^{-}\pi^{0}$        & $\phantom{0}6832 \pm 457$ & $15.71 \pm 0.16$         & $\phantom{0}2227 \pm 220$  & $15.93 \pm 0.29$\\
        $D_{s}^{-} \to K^{+}K^{-}\pi^{-}\pi^{0}$     & $23311           \pm 659$ & $10.58 \pm 0.05$         & $\phantom{0}7785 \pm 453$  & $10.39 \pm 0.08$\\
        $D_{s}^{-} \to K_{S}^{0}K^{-}\pi^{-}\pi^{+}$ & $\phantom{0}5269 \pm 282$ & $20.19 \pm 0.17$         & $\phantom{0}1662 \pm 217$  & $19.50 \pm 0.31$\\
        $D_{s}^{-} \to K_{S}^{0}K^{+}\pi^{-}\pi^{-}$ & $\phantom{0}8948 \pm 231$ & $21.63 \pm 0.09$         & $\phantom{0}3263 \pm 172$  & $21.29 \pm 0.15$\\
        $D_{s}^{-} \to \pi^{-}\pi^{-}\pi^{+}$        & $21909           \pm 776$ & $50.35 \pm 0.22$         & $\phantom{0}7511 \pm 393$  & $49.32 \pm 0.41$\\
        $D_{s}^{-} \to \pi^{-}\eta^{'}$              & $\phantom{0}4428 \pm 111$ & $19.00 \pm 0.08$         & $\phantom{0}1648 \pm 74\phantom{0}$ & $18.56 \pm 0.13$\\
        $D_{s}^{-} \to K^{-}\pi^{+}\pi^{-}$          & $10175           \pm 448$ & $47.19 \pm 0.32$         & $\phantom{0}4984 \pm 458$  & $45.66 \pm 0.59$\\
        \hline\hline
      \end{tabular}
    \end{center}
\end{table*}
        
\begin{table*}[htbp]
 \renewcommand\arraystretch{1.25}
  \caption{DT efficiencies ($\epsilon^{\rm DT}_{\alpha,\text{sig}, i}$) of each tag mode
    for the signal process $D^+_s\to f_0(980) e^+\nu_{e}$, $f_0(980)\to \pi^0\pi^0$ at (I) $\sqrt{s}= 4.178$~GeV,
    (II) $4.189-4.219$~GeV, and (III) $4.226$~GeV. Uncertainties are statistical only. These
    efficiencies do not include the following intermediate-state branching fractions for
    $K_S^0\to \pi^+\pi^-$, $\pi^0\to\gamma\gamma$, $\eta\to \gamma\gamma$, and $\eta^{\prime}\to\pi^+\pi^-\eta$. 
  }\label{tab:pi0_dtagEff2}
    \begin{center}
    \begin{tabular}{lccc}
\hline\hline
      Tag mode  & (I)$\epsilon^{\rm DT}_{\alpha,\text{sig}, i}(\%)$& (II)$\epsilon^{\rm DT}_{\alpha,\text{sig}, i}(\%)$&(III)$\epsilon^{\rm DT}_{\alpha,\text{sig}, i}(\%)$\\
      \hline
      $D^-_s\to K^0_{S}K^{-}$                      & $7.11\pm0.26$        & $6.64\pm0.13$         & $6.48\pm0.25$\\
      $D^-_s\to K^{+}K^{-}\pi^{-}$                 & $5.61\pm0.10$        & $5.24\pm0.05$         & $5.08\pm0.10$\\
      $D_{s}^{-} \to K_{S}^{0}K^{-}\pi^{0}$        & $1.93\pm0.13$        & $1.93\pm0.07$         & $1.87\pm0.14$\\
      $D_{s}^{-} \to K^{+}K^{-}\pi^{-}\pi^{0}$     & $1.43\pm0.05$        & $1.38\pm0.02$         & $1.41\pm0.05$\\
      $D_{s}^{-} \to K_{S}^{0}K^{-}\pi^{-}\pi^{+}$ & $2.74\pm0.20$        & $2.43\pm0.09$         & $2.38\pm0.18$\\
      $D_{s}^{-} \to K_{S}^{0}K^{+}\pi^{-}\pi^{-}$ & $2.48\pm0.15$        & $2.41\pm0.08$         & $2.40\pm0.14$\\
      $D_{s}^{-} \to \pi^{-}\pi^{-}\pi^{+}$        & $8.31\pm0.27$        & $7.73\pm0.13$         & $7.60\pm0.25$\\
      $D_{s}^{-} \to \pi^{-}\eta^{'}$
                                                   & $2.79\pm0.20$        & $2.51\pm0.10$         & $2.18\pm0.17$\\
      $D_{s}^{-} \to K^{-}\pi^{+}\pi^{-}$          & $7.05\pm0.32$        & $6.51\pm0.15$         & $6.55\pm0.29$\\
\hline\hline
    \end{tabular}
  \end{center}
\end{table*}

\begin{table*}[htbp]
   \renewcommand\arraystretch{1.25}
  \caption{DT efficiencies ($\epsilon^{\rm DT}_{\alpha,\text{sig}, i}$) of each tag mode
    for the signal process $D^+_s\to f_0(500) e^+\nu_{e}$, $f_0(500)\to \pi^0\pi^0$ at energy
    points, (I) $\sqrt{s}= 4.178$~GeV, (II) $4.189-4.219$~GeV, and (III)
    $4.226$~GeV. Uncertainties are statistical only.
    These efficiencies do not include the following intermediate-state branching
    fractions for $K_S^0\to \pi^+\pi^-$, $\pi^0\to\gamma\gamma$, $\eta\to \gamma\gamma$, and
    $\eta^{\prime}\to\pi^+\pi^-\eta$. 
  }\label{tab:sigma_dtagEff2}
  \begin{center}
    \begin{tabular}{lccc}
\hline\hline
      Tag mode  & (I)$\epsilon^{\rm DT}_{\alpha,\text{sig}, i}(\%)$& (II)$\epsilon^{\rm DT}_{\alpha,\text{sig}, i}(\%)$&(III)$\epsilon^{\rm DT}_{\alpha,\text{sig}, i}(\%)$\\
      \hline
      $D^-_s\to K^0_{S}K^{-}$                      & $4.34\pm0.20$        & $4.21\pm0.10$         & $3.64\pm0.19$\\
      $D^-_s\to K^{+}K^{-}\pi^{-}$                 & $3.61\pm0.08$        & $3.50\pm0.04$         & $3.30\pm0.08$\\
      $D_{s}^{-} \to K_{S}^{0}K^{-}\pi^{0}$        & $1.24\pm0.11$        & $1.15\pm0.05$         & $1.01\pm0.10$\\
      $D_{s}^{-} \to K^{+}K^{-}\pi^{-}\pi^{0}$     & $0.93\pm0.04$        & $0.94\pm0.02$         & $0.90\pm0.04$\\
      $D_{s}^{-} \to K_{S}^{0}K^{-}\pi^{-}\pi^{+}$ & $1.96\pm0.17$        & $1.70\pm0.08$         & $1.78\pm0.16$\\
      $D_{s}^{-} \to K_{S}^{0}K^{+}\pi^{-}\pi^{-}$ & $1.65\pm0.12$        & $1.79\pm0.06$         & $1.60\pm0.12$\\
      $D_{s}^{-} \to \pi^{-}\pi^{-}\pi^{+}$        & $5.30\pm0.22$        & $5.06\pm0.11$         & $4.47\pm0.20$\\
      $D_{s}^{-} \to \pi^{-}\eta^{'}$
                                                   & $1.73\pm0.16$        & $1.54\pm0.08$         & $1.51\pm0.15$\\
      $D_{s}^{-} \to K^{-}\pi^{+}\pi^{-}$          & $4.13\pm0.24$        & $4.36\pm0.13$         & $3.96\pm0.24$\\
\hline\hline
    \end{tabular}
  \end{center}
\end{table*}

\begin{table*}[htbp]
 \renewcommand\arraystretch{1.25}
  \caption{DT efficiencies ($\epsilon^{\rm DT}_{\alpha,\text{sig}, i}$) of each tag mode
    for the signal process $D^+_s\to K_{S}^{0}K_{S}^{0} e^+\nu_{e}$ at (I) $\sqrt{s}= 4.178$~GeV, (II) $4.189-4.219$~GeV, and (III)
    $4.226$~GeV. Uncertainties are statistical only. These efficiencies do not include the following intermediate-state branching
    fractions for $K_S^0\to \pi^+\pi^-$, $\pi^0\to\gamma\gamma$, $\eta\to \gamma\gamma$, and $\eta^{\prime}\to\pi^+\pi^-\eta$. 
  }\label{tab:KsKs_dtagEff2}
  \begin{center}
    \begin{tabular}{lccc}
\hline\hline
      Tag mode  & (I)$\epsilon^{\rm DT}_{\alpha,\text{sig}, i}(\%)$& (II)$\epsilon^{\rm DT}_{\alpha,\text{sig}, i}(\%)$&(III)$\epsilon^{\rm DT}_{\alpha,\text{sig}, i}(\%)$\\
      \hline
      $D^-_s\to K^0_{S}K^{-}$                      & $4.87\pm0.22$        & $4.51\pm0.10$         & $4.85\pm0.21$ \\
      $D^-_s\to K^{+}K^{-}\pi^{-}$                 & $3.84\pm0.08$        & $3.65\pm0.04$         & $3.78\pm0.08$ \\
      $D_{s}^{-} \to K_{S}^{0}K^{-}\pi^{0}$        & $1.41\pm0.12$        & $1.37\pm0.06$         & $1.31\pm0.11$ \\
      $D_{s}^{-} \to K^{+}K^{-}\pi^{-}\pi^{0}$     & $1.00\pm0.04$        & $1.03\pm0.02$         & $0.97\pm0.04$ \\
      $D_{s}^{-} \to K_{S}^{0}K^{-}\pi^{-}\pi^{+}$ & $1.25\pm0.14$        & $1.04\pm0.06$         & $1.01\pm0.12$ \\
      $D_{s}^{-} \to K_{S}^{0}K^{+}\pi^{-}\pi^{-}$ & $1.22\pm0.10$        & $1.23\pm0.05$         & $1.30\pm0.11$ \\
      $D_{s}^{-} \to \pi^{-}\pi^{-}\pi^{+}$        & $5.56\pm0.23$        & $5.09\pm0.11$         & $4.78\pm0.21$ \\
      $D_{s}^{-} \to \pi^{-}\eta^{'}$
                                                   & $1.42\pm0.15$        & $1.69\pm0.08$         & $1.54\pm0.15$ \\
      $D_{s}^{-} \to K^{-}\pi^{+}\pi^{-}$          & $4.44\pm0.25$        & $4.32\pm0.12$         & $4.30\pm0.25$ \\
      \hline\hline
    \end{tabular}
  \end{center}
\end{table*}